# 75 Years of Matter Wave: Louis de Broglie and Renaissance of the Causally Complete Knowledge

A.P. KIRILYUK*

Institute of Metal Physics, Kiev, Ukraine 03142

> *Mince, discret, secret, Louis de Broglie se tient toujours à l'entrée de la route qu'il ouvrit et, avec un sourire courtois de gentilhomme, regarde l'avenir passer.*
>
> Maurice Druon, Préface à *Louis de Broglie* (1973) [16]

ABSTRACT. A physically real wave associated with any moving particle and travelling in a surrounding material medium was introduced by Louis de Broglie in a series of short notes in 1923 and in a more complete form in his thesis defended in Paris on the 25th November 1924. This result, recognised by the Nobel Prize in 1929, gave rise to a major direction of "new physics" known today as "quantum mechanics". However, although such notions as "de Broglie wavelength" and "wave-particle duality" form the basis of the standard quantum theory, it actually only takes for granted (postulates) the formula for the particle wavelength (similar to other its formally postulated "results") and totally ignores the underlying causal, realistic and physically transparent picture of wave-particle dynamics outlined by Louis de Broglie in his thesis and further considerably developed in his later works, in the form of "double solution" and "hidden thermodynamics" theory (the first of them is only schematically outlined in de Broglie's "pilot-wave" description, or "Bohmian mechanics", often called today "causal de Broglie-Bohm theory"). A price to pay for such rough deviation from the original de Broglian realism and consistency involves fundamental physics domination by purely abstract and mechanistically simplified schemes of formal symbols and rules that have led to a deep knowledge impasse justly described as "the end of science". However, a new, independent approach of "quantum field mechanics" (quant-ph/9902015, 16, physics/0401164) created recently within the "universal science of complexity" and related "dynamic redundance paradigm" (physics/9806002) provides many-sided confirmation and natural completion of de Broglie's "nonlinear wave mechanics", eliminating all its "difficult points" and reconstituting the causally complete, totally consistent and intrinsically unified picture of the real, complex micro-world dynamics directly extendible to all higher levels of unreduced world complexity.

---

*Address for correspondence: Post Box 115, Kiev-30, Ukraine 01030. E-mail: kiril@metfiz.freenet.kiev.ua.

## CONTENTS

---

*…les progrès de la Science considérés indépendamment de leurs applications ont toujours résulté d'efforts pour "mieux comprendre". Le désir de comprendre a été à l'origine de toutes ses réussites. Aussi suis-je aujourd'hui amené à penser qu'il convient d'être très réservé en face de l'affirmation, si souvent répétée par de nombreux physiciens depuis une quarantaine d'années, suivant laquelle les phénomènes de transition quantique transcenderaient, pour employer un mot de Niels Bohr, toute description en termes d'espace et de temps et seraient par suite définitivement incompréhensibles. Il me paraît plus naturel et plus conforme aux idées qui ont toujours heureusement orienté la recherche scientifique de supposer que les transitions quantiques pourront un jour être interprétées, peut-être à l'aide de moyens analytiques dont nous ne disposons pas encore, comme des processus très rapides, mais en principe descriptibles en termes d'espace et de temps, analogues à ces passages brusques d'un cycle limite à un autre que l'on rencontre très fréquemment dans l'étude des phénomènes mécaniques et électromagnétiques non linéaires.*

    *Plus généralement je dirai que, quand un processus physique nous paraît impossible à comprendre et à représenter, nous devons toujours penser que de nouveaux et vigoureux efforts intellectuels nous permettrons un jour de le comprendre et de le représenter.*

    Louis de Broglie, *Les idées qui me guident dans mes recherches* (1965), In [40]

---

# 1. Creation of causal wave mechanics by Louis de Broglie, its imitation by the abstract (standard) quantum mechanics, and deceptive results of the twentieth century "scientific revolution"

Since a truly revolutionary discovery changing qualitatively fundamental scientific concepts inevitably enters in antagonistic contradiction with the "established" views, it often appears in an unexpected fashion, on a "minor" by-way and practically always needs much efforts and time for its full understanding, even after the formal "acknowledgement" by the scholar system of "well-established knowledge". A good illustration of this rule is provided by the destiny of revolutionary discovery of the wave properties of particles (and thus "wave-particle duality") initiated by Louis de Broglie in 1923-24, then successfully confirmed and formally recognised, but remaining profoundly misunderstood and incomplete until the now fading end of the twentieth century.

The "wave associated with a (moving) material particle (mobile)" was introduced by Louis de Broglie in three short notes presented to the French Academy of Sciences in September-October 1923 [1-3] (see also ref. [4] for an outline in English) and in complete form in his historical thesis defended on the 25th November 1924 in Paris [5] and preceded by some development of the primary ideas [6-8]. The thesis was defended with excellence according to the report signed by such prominent scientists as Jean Perrin and Paul Langevin [5]. Most remarkable, however, is the fact that the work of a young researcher contains a totally new, physically transparent and mathematically confirmed outlook on the origin of matter and dynamic behaviour of its elementary constituents, initiating a formidable *synthesis* of the emerging "new physics" ideas (relativity and "theory of quanta") and unreduced conceptual *realism* of the "old", "classical" physics (classical mechanics, optics, and thermodynamics).

De Broglie shows in his thesis that "relativistic equivalence of mass and energy", combined with the fundamental (Planckian) quantization of any form of energy, imposes the existence of internal oscillation process within each (elementary) material particle, which should naturally "develop" itself into an extended, physically real wave when the particle moves in a surrounding material continuum, so that the internal, localised oscillation of the particle-mobile should remain permanently in phase with the travelling wave oscillations. This result, designated as the "phase accord theorem" and based on the profoundly causal physical picture of the *emerging* wave-particle duality, leads to the famous mathematical expression for the length, $\lambda$, of that accompanying, or "guiding", matter wave, if one demands that the *different* relativistic transformations of the frequency (or the related time flow) of the moving localised oscillation and particle mass do not violate coherence between them (i. e. actually the physical integrity of a compound object):

$$\lambda = \frac{h}{mv} , \qquad (1)$$

where $m$ is the (total, relativistic) mass and $v$ is the *classical* velocity of the particle. The thoroughly realistic, physically transparent, but non-trivial, and "everywhere continuous" combination of the hypothesis of quanta and canonical relativity is further confirmed by de Broglie using independent and equally causal unification of the fundamental principles of classical mechanics and optics (the principles of Maupertuis and Fermat, respectively) assisted by important additional involvement of classical thermodynamics and relativistic electrodynamics, so that the final expression of the physical wave-particle duality, eq. (1), emerges from the whole unified picture only on page 111 of the original thesis edition [5] (ending at page 128):

"Si les vitesses sont assez faibles pour permettre de négliger les termes de Relativité, la longueur d'onde liée au mouvement d'une molécule dont la vitesse est $v$, sera:

$$\lambda = \frac{\frac{c}{\beta}}{\frac{m_0 c^2}{h}} = \frac{h}{m_0 v} \ldots\text{''}$$

It is therefore quite evident from the original foundation of wave-particle duality and derivation of its mathematical expression in de Broglie's work that what he proposes is not only a "formula", though "experimentally verifiable" and "practically useful", but a qualitatively new level of understanding of the physical reality and its detailed dynamical structure, produced by a non-trivial union of various fundamental approaches and principles, which is absent in the "old", classical physics, but at the same time naturally, continuously emerges from it within the causal, realistic, and thus universal, logic of any truly scientific knowledge-understanding. It is this *reality-based unification* of the old-physics results that produces the genuine new-physics basis, in its original formulation by de Broglie.

However, the intrinsic, *extended causality* of the original approach of Louis de Broglie (*mécanique ondulatoire* [9]), profoundly related to the special "humanitarian" quality of his thinking and aimed at the *complete understanding* of physical reality (see e. g. [10]), was then quickly lost within a simplified, purely abstract adaptation of his results in the "standard" (since 1927) quantum mechanics, promoted especially by N. Bohr and his followers ("Copenhagen interpretation") and constituting a part of the general mechanistic and totally formal approach of "mathematical physics" which, in order to survive, *should* be detached from the physical reality it tries to imitate. Specifically, the scholar quantum mechanics borrows from de Broglie's "undulatory (wave) mechanics" only the famous formula for the "de Broglie wavelength" of a moving particle, eq. (1), while completely ignoring the fundamentally important, reality-based picture of dualistic dynamics underlying it, used by de Broglie for that formula derivation, and actually constituting, as we shall see, a part of the underlying *complex interaction dynamics* [11-14]. Even apart from the well-known physical incompleteness and logical contradictions of the "mathematical" quantum mechanics, readily recognised by the canonical sources (see e. g. [15]), the wavelength expression contains a strange, "impossible" mixture of classical properties of a particle (velocity and mass), its "quantized" (Planck's constant) and "wave" (wavelength) behaviour that can be but superficially hidden behind the formal "postulates" of the canonical, "math-physical" approach.[1]

The well-organised, "collective" and "ideological", rejection of de Broglian realism by the "international community" of abstract-minded "wunderkinds", who did not need any reality at all for their cabalistic exercises, had taken the especially strong form of a "coup d'état" [10] during the Fifth Solvay Congress, in 1927, after which the unreduced causality of the original wave-particle duality quickly became extinct by the strict *subjective* exclusion of *any* causal approach from quantum mechanics and (new) fundamental physics in general, despite objections from a number of prominent physicists (like those of E. Schrödinger and A. Einstein) and high official recognition of de Broglie's results by the Nobel Prize (in 1929) and other rewards. Because of this sharp divergence with the "generally accepted values" of the resulting totally formalised science of the twentieth century, relying on the "well-established" (especially since Solvay Congresses) method of "truth by vote", Louis de Broglie was obliged to interrupt, for the next 25 years, his solitary search for the causally complete understanding of the unique and unified reality and, contrary to his natural inclination, devoted his efforts in fundamental science mainly to teaching of the canonical "new physics", combining it with a background investigation of particular problems [16] (the details of dramatic development of de Broglie's life and work can be found in the exciting, but scientifically rigorous story of ref. [10]).

---

[1] Thus, the dualistic, complex dynamics of the physically real "wave-particle" in de Broglian approach, underlying eq. (1), is replaced in the canonical theory by artificial introduction of abstract "momentum operator" (opening a whole series of equally formal "operators" and other ever more abstract constructions), which is accompanied by a *postulated* expression for its formal, mathematical "action" on an abstract "state vector" [15], simplistically imitating the real, dynamic wave-particle duality with the help of de Broglie's formula taken for granted, despite its causal derivation in the thesis of 1924.

However, such is the power of the unreduced realism and causally complete knowledge that after 25 years of forced suspension Louis de Broglie finds the new élan for development of his original ideas and since 1952, at his age of 60 (and until his retirement around 1975) produces an impressive number of fundamentally important new results, considerably completing and amplifying the causal picture of wave-particle duality and concentrated around the famous "double solution" [17-19] and related "hidden thermodynamics of the isolated particle" [20-23] (see also refs. [16,24-27]).[2]

In the meanwhile, the canonical, "math-physical" quantum mechanics has remained at the same basic level quickly attained just after its creation during the twenties, but having never changed since then despite the growing number of efforts using ever more sophisticated abstractions of the canonical symbolism. During several tens of years after its establishment, this stagnation of the canonical quantum theory was rather efficiently masked by its appearing new applications in atomic, then particle physics, chemistry, solid state physics, etc., even though it was quite clear that they have never brought anything new to the (absent) fundamental problem solution. However, contrary to a popular belief, the number of really meaningful, original applications of even a very fundamental theory is not infinite, and as the stock of "interesting problems" in quantum mechanics was becoming the more and more exhausted, in the second half of the century (not to mention a vanishingly small proportion of really useful solutions obtained), its conceptual incompleteness and glaring contradictions have come again to the foreground of fundamental science. This new and ever growing interest in the foundation problems is closely related to similar "explicative" difficulties in other branches of fundamental physics stemming from the same root, such as "field theory", "particle physics", (quantum) gravitation, and cosmology, that remain as causally deficient and contradictory as the canonical "quantum theory", but also basically separated among them, despite the common nature of the objects of study and a large number of efforts put to attain unification (such as various abstract schemes of "unification of interaction forces" within the canonical field theory). Practical importance of those fundamental problems also grows in an unexpected way, since the technical power of science has experienced a dramatic, "revolutionary" increase in the same period, and now all that huge and sophisticated machinery finds itself in the impasse of absent sensible, understandable and controlled, applications leading towards true discoveries (their absence proves the conclusion) and involves instead quite real dangers created by the remaining purely empirical, "trial-and-error" use of that immense, artificial, and increasingly destructive power, now systematically exceeding the "self-organised" limits of natural phenomena. Fruitless "shooting in all directions" from super-powerful "guns" realised by super-expensive super-accelerators provides a characteristic example of that scholar science tendency and real dangers it involves [11,14].

---

[2] It is important to emphasize here the fundamental difference between the unreduced, *essentially nonlinear* version of de Broglie's undulatory mechanics, expressed by the double solution concept, and its simplified imitation within so-called "Bohmian mechanics", which is often referred to as "de Broglie-Bohm theory" and characterised as "causal approach in quantum mechanics" by its adherents (see e. g. [28-30]). In reality, however, the Bohmian mechanics is not more causal than any other "interpretation" of the conventional quantum mechanics: without changing the formal, postulated character of the theory and finding the true, reality-based solution of the ensuing "quantum mysteries", each "interpretation" tries to accentuate one or another aspect of the canonical "quantum enigmas" by mechanically playing with the same *abstract* symbols and trying to produce arbitrary, pseudo-philosophical guesses about what the resulting formal constructions *could* mean. The reference to de Broglie in the name of "de Broglie-Bohm interpretation" is the acknowledgement of the undeniable fact that its main results, before being "rediscovered" by D. Bohm in 1952 [31], have been obtained by Louis de Broglie 25 years earlier, yet in 1927, under the name of "pilot-wave description" [9] (and that explains why Bohm's paper of 1952 gave a subjective "impact" to de Broglie, contributing to his motivation to restart the search for the unreduced double solution [10]). However, de Broglie considered that kind of description only as an *intentionally* reduced mathematical *scheme*, obtained by simple algebraic (identical) transformation of the canonical Schrödinger formalism and formally reproducing but an external part of the complete double solution that escaped the unreduced mathematical presentation, but remained nevertheless the main object of the "veritable basic ideas of the undulatory mechanics" [25] (see also [10,16-19,24,26,27]). Whereas the unreduced de Broglie theory is totally devoted to the search for the adequate mathematical representation of the physically complete, realistic understanding of the dynamical micro-world structure, the adherents of "Bohmian mechanics" persist in replacing the true causality by the same mechanistic, mathematically based manipulations with symbols, but presented as a "physically consistent", "causal" version of quantum mechanics (cf. [32]), while the unreduced double solution concept underlying the original "pilot-wave" scheme is thoroughly excluded from their work. However, one can neither fool nature, nor bribe the truth (contrary to the "general public", science publishers and "administrations"), and the unresolved contradictions of the canonical quantum mechanics inevitably come out in its Bohmian interpretation in the form of unexplained origin of quantum objects and their properties, such as *postulated* "particles" and (special) "waves", their permanent, "causeless" and unpredictable transformation into one another, and the intrinsically probabilistic character of all the occurring processes [11].

Those non-trivial, and ever growing, "difficulties" of the canonical fundamental science, readily acknowledged by many its prominent creators (see e. g. [33-38]) and now taking the scale of the "end of science" [39] (see also [11]), are in reality but inevitable payment for the deviation from the road of causally complete, unreduced understanding of reality that determined the development of "classical" science, but then was deliberately rejected by the "international scientific community" dominated by the inherent "mathematical physicists", with their characteristic inclination for absolute power of low-level, abstract calculations, "justified" by obscure, artificially mystified speculations under superficial slogans of *completely* "new" physics [40]. The "road of truth" was maintained almost exclusively by the singular efforts of Louis de Broglie and his followers. Indeed, there is no wonder that the canonical fundamental science cannot understand the "veritable", ultimate origin of the observed dynamical structure of the world taking into account the fact that the dominating over-simplified, abstract "models" are not intended for, and even opposed to, such unrestricted, intrinsically complete understanding of nature/reality. There is no wonder either that becoming the more useless the more their mechanistic sophistication and divergence from reality grow with time, the abstract constructions of the canonical theory provoke the overflowing mistrust and disgust of the "general public", younger generations and society in the whole. Such attitude is inevitably, though wrongly, extended to science in general, to any kind of scientifically ordered knowledge, since the canonical science, while readily describing the growing "mistrust" in full detail (see [11] for references), puts at the same time all its well-organised efforts to suppression of dissemination of *another*, realistic and fully adequate, creative and always developing kind of scientific knowledge, which is also much more rigorous (i. e. simply honest), both mathematically and physically, than canonical imitations always based on incorrect approximation within a version of "perturbation theory", but also has, contrary to them, a "human-friendly", reality-based and transparent character. It is difficult to overestimate the losses produced by such shameless substitution from the part of the educated mediocrity thoroughly organised in a network of "relations" and mutual "services" permeating the whole "developed" society. It is also clear that such huge deviation from the road of creation, involving in reality all aspects of life, cannot give anything else than the (already observed) rapid degradation of the quality of human living (always based on the quality of human thinking) down to a series of more pronounced falls, or "catastrophes", so clearly "felt" within various modern "prophecies" and objectively provoked precisely by that over-simplified, mechanistic approach to the intrinsically complex natural phenomena. This link between causality in science, creativity of thinking and "general" development of society, as well as the "critical" character it acquires today, was unambiguously recognised and emphasised by Louis de Broglie who devoted a large part of his activity and a whole series of publications to the fight against growing dangers of unlimited mechanistic simplification in the content and organisation of science (see e. g. [41] and other references in [10,16]).

This conclusion about the true, unfortunately suppressed, and false, artificially imposed, directions of knowledge development obtains decisive confirmation from the recently appeared completion of the unreduced double solution idea, realised within the independent, new concept of universal dynamic complexity [11] and produced within the same type of reality-based, intrinsically causal approach as de Broglie's "nonlinear wave mechanics" (contrary to many recent imitations around "science of complexity" à la Prigogine or Santa Fe & Co. fundamentally deficient by the same limitations of "one-dimensional" reduction of reality to positivistic classification of symbols, cf. [42]).

## 2. Universal science of complexity as the natural completion of original ideas of Louis de Broglie

### 2.1. The unreduced dynamic complexity and its imitations

The unreduced, intrinsically universal dynamic complexity naturally emerges in the unrestricted analysis of a generic interaction process within a system of primal (or arbitrary) entities [11] where no usual perturbational "cutting" of "inconvenient" ("non-separable") entanglement of physically real interaction participants is performed. The approach is realised with the help of generalised, universally nonperturbative version of the well-known "effective (optical) potential" method (see e. g. [43,44]) that reveals the natural structure of a problem, with the "inseparable, but integrable" part, giving the source of true chaoticity and complexity and determining the current level of unreduced structure emergence in the course of interaction development, and the related "secondary" interaction part, which is not rejected, however, but is further considered within the same analysis to give the next level of the emerging real structure, etc. (so that one finally obtains the *dynamically fractal* structure of the forming compound entity as the truly *complete solution* of a problem, representing the essential, causal extension of purely mathematical, regular imitations of the natural fractality).

The key result of this unrestricted, universally non-perturbative analysis of a generic system with interaction, directly related to causal completion of the nonlinear undulatory mechanics, is the naturally emerging phenomenon of *dynamic redundance (or multivaluedness)* of the obtained solutions and real structures they describe. Physically, dynamic redundance originates from the fact that each of the interacting entities exists, by definition, in the *same* reality as the result of their interactive *entanglement*, but the unrestricted character of a generic interaction process means that the number of combinations of "everything with everything" within the compound system will always be much greater than the available "space" in reality previously occupied by non-interacting partners. Therefore, the unreduced interaction analysis by the above "method of effective dynamical functions" shows that the truly complete problem solution has many "versions", each of them being "complete" in the ordinary sense (i. e. totally realistic) and therefore *incompatible* with any other, *equally real*, version. The multiplicity of these component solutions, called *realisations* of the system (problem), is easily obtained from the simple analysis of the mentioned "pseudo-integrable" part of the problem, whereas its general nonintegrability manifests itself in a relatively weak, but irreducible, link of this level of structure formation to the next level of finer splitting into another redundant set of solution-realisations, etc.

The only possible issue from the "conflict" among realisations, at each level of structure formation, is their permanent change, where the system unceasingly and autonomously "switches" from one of its realisation to another. Since all the elementary realisations are strictly equal in their "rights for existence", one cannot predict which realisation the system will take next, and this gives us the *causal*, purely *dynamic* origin of *naturally emerging randomness* and related *probability* in a generic system with interaction.[3] Namely, it is clear that the probability, $\alpha_r$, of emergence of a certain, $r$-th, elementary system realisation, constituting causal extension of the canonical, *postulated* notion of *event* in the mathematical "theory of probability", is equal to $1/N_\Re$, where $N_\Re$ is the total number of realisations of the current level, and now it is a *dynamically determined* quantity that can be obtained *analytically* from the dynamic equation describing the interaction process. In a general case, the elementary realisations can be dynamically grouped into dense agglomerates (or dynamical tendencies), so that individual elementary realisations are not resolved experimentally within groups that play the role of actual, "combined" system realisations whose probabilities are not equal any more:

$$\alpha_r(N_r) = \frac{N_r}{N_\Re} \quad \left(N_r = 1, ..., N_\Re; \sum_r N_r = N_\Re\right), \quad \sum_r \alpha_r = 1, \tag{2}$$

where $N_r$ is the number of elementary realisations within the $r$-th combined (actual) realisation ($N_r$ is also determined dynamically), and the sum is taken over all such actually observed system realisations. If, for a given system and level of its structure formation, all $N_r$ (and thus $\alpha_r$) tend to be comparable among them (and usually small, $\alpha_r \ll 1$), we have the limiting regime of *uniform chaos* characterised by a high observed irregularity in the system behaviour, while the opposite case of essentially differing, relatively large $N_r$ ($\alpha_r \sim 1$) corresponds to the characteristic regime of (distinct) structure formation (dynamically multivalued *"self-organisation"*, or *"self-organised criticality"*, SOC) and suppressed external irregularity (even though it is *always* present *within* the observed quasi-regular patterns and remains indispensable for their very existence). Both limiting regimes of causal randomness, or *(dynamical) chaos* (thus rigorously defined), as well as all possible intermediate situations occurring at all levels of interaction development process, account *completely* for the *whole* diversity of the observed structures of the world and their dynamics [11].

---

[3] This result of the unreduced science of complexity is essentially different from its mechanistic imitation in the canonical, fundamentally single-valued description of "complex behaviour" (conventional "chaos theory") that invokes already existing, externally imposed "attractor components" ambiguously *coexisting* in an *abstract* mathematical "space" and leading to empirically observed "multi-mode behaviour". As a result, the canonical "science of complexity" does not see either the true origin, or universality of unreduced, *dynamic* multivaluedness and the ensuing *permanently*, autonomously maintained causal randomness, in the form of *unceasing* change of *incompatible* (rather than "coexisting") realisations. Similar to any other application of the effectively one-dimensional paradigm of the canonical science, this basically inconsistent "theory of chaos" without chaos (true, dynamic randomness) tends to consider an isolated system taking *only one* dynamical regime, or "mode", once and forever, even if "in principle" there are many of them, "somewhere" (see [11] for more details).

The quantitative measure of *dynamic complexity*, *C*, providing the unique, unifying property of any part of that diversity of chaotic behaviour, is provided by any growing function of either realisation number, or rate of their change, equal to zero for the unrealistic case of only one system realisation (for example, $C \propto \ln(N_\Re)$), where that fatally simplified case constitutes the *only* situation considered within the *whole* canonical, *always dynamically single-valued* (or *unitary*) science. This huge reduction of the canonical science paradigm is the *rigorously specified* reason for all its "irresolvable difficulties" (like perturbation series divergence) and persisting "mysteries" (like those from the canonical quantum mechanics, see below). It is related to conventional perturbative "cutting" of essential interaction links transforming any real problem from its original, "nonintegrable" quality to a deceptively convenient, "integrable" (or "separable", "exactly solvable") one, but simultaneously "killing" the universal mechanism of autonomous structure formation through the naturally forming hierarchy of self-amplifying "feedback loops" of "nonintegrable", real interaction process [11-14]. This unreduced, *dynamically emerging* hierarchy of self-amplifying interaction loops provides also causal extension of the notion of *nonlinearity* and shows that *any real* interaction is an intrinsically nonlinear process, contrary to ambiguous ideas about "nonlinearity" in the canonical science (since the latter deals always with effectively one-dimensional, and thus linear, processes, it tends to imitate nonlinearity by a "curved", "folded", or "pointed" line, always remaining, however, a one-dimensional, and artificially produced, construction). The extended, *essential, or dynamic, nonlinearity* appears mathematically as unreduced effective potential dependence on the eigen-solutions to be found, contrary to perturbative equations of the canonical theory, where solutions to be found *cannot* occur *within* the expression of "interaction operator" acting upon them (the property that leads "self-consistently" to formal canonical theorems about the "uniqueness of solution"), and therefore the naturally emerging nonlinearity of "operators" is ruthlessly cut by invariably perturbative, one-dimensional logic of the canonical science [11-14] (see also the next section).

It is easy to understand also that unpredictably emerging realisations form the physical structure (or "texture") of *space* at the corresponding level of complex dynamics, while the facts of their appearance (i. e. the sequence of causally specified *events*) provide the *unceasing* flow of *time*. This causally obtained, physically "tangible" space is naturally *discrete*, while equally real time is naturally *irreversible* (because of the causal randomness of event sequence), but is *not* a tangible, "material" entity and therefore *cannot* be considered as another physical "dimension" and "unified" with the real, tangible space dimensions (contrary to abstract simplification of the canonical relativity). One can see that causal time and space, together with the whole arborescence of dynamic complexity, include the hierarchical structure of levels (where the lowest level plays the role of the "embedding", most fundamental space and time, see the next section) and possess the natural property of relativity, resulting directly from their dynamical, causal origin [11-14] and providing realistic extension of the purely formal, *postulated* interpretation of Lorentz-Poincaré relations of the canonical relativity.

The intrinsic irreversibility of the causal time flow, directly related to the dynamic redundance phenomenon, provides also causal understanding of *nonunitary* character of any dynamical system existence/evolution as unceasing, qualitatively inhomogeneous alternation of (regular) motion within realisation and highly irregular, unpredictable transition to the next realisation, as opposed to *intrinsically unitary* (i. e. qualitatively *uniform*) character of any dynamic (interaction) process considered within the dramatically limited framework of the canonical science, with its *single* realisation, arbitrarily fixed "once and forever". In that way one obtains also the extended, universal understanding of the internal structure, and the physical sense itself, of any real "interaction" which is represented by that permanent chaotic change of qualitatively different phases of "strong interaction" between closely entangled interaction partners (within each system realisation) and "weak interaction" within a quasi-free, irregular motion of transiently disentangled partners (during chaotic transitions between realisations), as opposed to the canonical idea of qualitatively smooth (actually, averaged), regular distribution of "interaction magnitude".

Since the same, universal mechanism of dynamical system splitting into incompatible realisations occurs at each emerging level of structure formation process, the resulting *dynamical fractal* that represents the whole multi-level arborescence of system dynamics, has the intrinsically *probabilistic* and *self-developing* character, which means that its "branches" permanently *grow* in an unpredictable (but partially, probabilistically ordered) fashion and in this way the system *autonomously* finds the (irregular) route of its intrinsic development, being driven by the same interaction. This universal behaviour of a system with interaction provides the unified causal understanding of the observed property of *dynamic adaptability*, or *creativity* (being another expression of dynamic nonlinearity) and essential extension of the canonical notion of fractal (Mandelbrot *et al.*), devoid of any *intrinsic*, dynamical randomness and adaptable development (creativity). In particular, our fundamental dynamical fractal (of a problem/system), representing the unified, truly complete *general solution* of a problem, can equally well describe, at different levels of its structure, both "apparently fractal", "fuzzy" objects (like "coast line", or "crystallisation patterns") and "apparently non-fractal", "distinct" objects (like a pebble reposing at the "coast line", or an egg). We can say that *any* real dynamical system can be considered as being *alive*, in a well-defined sense, since it shows such basic features of a "living creature", now causally understood, as autonomous dynamic creativity, adaptability, and growth towards the more and more involved structure elements (needless to say, the "real life", in the narrow sense, appears starting from a certain, high enough level of complexity including the ability to maintain the existence of a well-determined *species* of systems, by way of *coding, or informational, reproduction* [11]).

The general "direction of life", or fractal structure development, is well defined by the *growth* of complexity of its tangible, unfolded structure representing the causal, unified extension of the canonical *entropy*, while the dual form of complexity, called *(dynamic) information* and characterising the "folded" (or hidden) "stock" of the "future" complexity-entropy at the *start* of interaction process, correspondingly *decreases*, being directly *transformed* into dynamic entropy, so that the *total dynamic complexity*, equal to the sum of both its forms, remains unchanged. This universal law of *conservation (or symmetry) of complexity* can be logically justified within the universal science of complexity, and being equivalent to the causally complete forms of *all* known "conservation laws", "variational" and various other "principles" (like the "second law of thermodynamics" or "principle of relativity"), it is confirmed by the *totality* of existing experimental observations [11].[4] Providing the unified law, and criterion, of (progressive) development/evolution of any system and the whole world, the universal symmetry of complexity avoids any mechanistic "symmetry of identical configurations" of the canonical science and *dynamically* produces symmetric, but *irregular, variable* structures (it is therefore *always exact* and never "broken", contrary to any unitary symmetry).

Similar to fractality, other complexity-imitating versions of the unitary science are basically deficient and therefore can only mechanically repeat certain external, and irreducibly separated, properties, or "signatures", of complex behaviour (see e. g. [11,42]). In particular, "complex" systems considered within the canonical science are always some "special" systems, or configurations, or regimes of behaviour opposed, explicitly or implicitly, to other, "non-complex" (regular and simply structured) systems, without any clear criterion of distinction between the two (let alone other ambiguously intermixed notions, like "nonlinearity", nonintegrability, etc.), whereas the universal science of complexity demonstrates, simply due to the unreduced analysis of an arbitrary driving interaction, that *any* really existing system is a dynamically complex one, which means that it has many clearly defined (and consistently derived), permanently and chaotically changing realisations. Characteristic, and most popular, imitations of the single-valued "science of complexity" are provided by the concepts of "chaos" that should result from "exponentially diverging trajectories" and "self-organisation" (or "synergetics") considering only single-valued, "exact" solutions, both of these "concepts" being deduced within the explicitly perturbative analysis, illegally (formally) applied to the essentially nonperturbative case of complex dynamics. General and particular inconsistencies, resulting from such evident manipulations, immediately show themselves, and it becomes clear, in particular, that system states can diverge only according to a power law and not exponentially (besides the evident fact that *any* regular dependence cannot serve as a *fundamental* source of randomness), whereas any "self-organised" structure can maintain its shape only due to the internal chaotic change of (sufficiently close) realisations organised in a fractal, "nonintegrable" hierarchy (see [11] for more details).

It is not difficult to see that it is precisely the characteristic features of the unreduced dynamic complexity, distinguishing it from the unitary science imitations ("signatures of complexity"), that naturally show the inherent properties of quantum behaviour, such as the intrinsic source of true randomness, dynamic duality between the "localised", regular system state within each realisation and its delocalised, irregular state during chaotic "jumps" between realisations, as well as the nonlinear dynamic "reduction" (or "collapse") of the latter, delocalised state when it is transformed into each localised realisation, determining system "configuration" under the influence of driving interaction. That is why it is especially important to avoid imitations of the unitary "science of complexity" while considering the relation between dynamic complexity and quantum behaviour, the latter indeed representing, as we shall see, a situation with pronounced, irreducible (though finally standard) manifestations of the dynamic redundance phenomenon within a system of interacting entities (so that the very persistence of "quantum mysteries" is a result of invariable and total neglect of unreduced complexity within the effectively one-dimensional approach of the canonical science).

Equally important is the *universal* character of characteristic features of the unreduced complex behaviour leading to rigorously derived generalisation of "purely quantum" properties to *any* system behaviour represented now in its complete, irreducibly complex (dynamically multivalued) form, which permits us to eliminate explicit or hidden inconsistencies of the canonical "classical" science (also single-valued and unitary) and extend it to the ultimately universal and truly adequate knowledge *exactly* reproducing the observed *unified diversity* of the *whole* existing world [11]. Without any coincidence, the same tendency towards ever growing unification of the basic principles of physics is clearly expressed in the work of Louis de Broglie [10,11,16,25-27].

---

[4] This causal, reality-based, and universal extension of the notions of entropy and information permits one to escape various ambiguous speculations around these quantities in the unitary science (see [11] for more detail). Being limited to a single, qualitatively uniform realisation, canonical science cannot consistently introduce any useful concept of entropy or information, and tries to imitate them by empirically counting the "observed" (or imaginary) objects and putting the found number under the sign of logarithm chosen by mechanical adjustment to observation results for simple "model systems". The single-valued paradigm of the canonical science cannot provide any consistent idea about qualitative development (change) and thus the difference (transformation) between information and entropy, and therefore it repeatedly and arbitrarily deforms and confuses them, totally missing the law of total complexity conservation.

## 2.2. Causally complete wave mechanics (quantum field mechanics) as complex interaction dynamics

______________________________________________________________

*En conclusion, je pense que mes idées primitives, telles que je les ai reprises et développées dans ces dernières années, permettent de comprendre la véritable nature de la coexistence des ondes et des particules dont la Mécanique quantique usuelle et ses prolongements ne nous donnent qu'une vue statistique exacte sans nous en révéler la véritable nature. Le postulat de l'accord des phases nous apprend, en effet, qu'il existe une Dynamique à* masse propre variable *qui est sous-jacente à toute propagation d'ondes, même quand celle-ci s'effectue en dehors de l'approximation de l'optique géométrique. Et je crois que c'est là ce que la Mécanique quantique actuelle n'a pas su voir.*

Louis de Broglie, "Sur les véritables idées de base de la Mécanique ondulatoire",
C.R. Acad. Sc. (série B) **277** (1973) 71 [25]

______________________________________________________________

The fundamental dynamic multivaluedness was first revealed in the study of charged particle scattering in crystals performed within the Schrödinger formalism [45], and then was directly generalised to description of the true quantum chaos in a generic Hamiltonian quantum system [46] and quantum measurement process [47] (see also ref. [11]). Each of these situations is described as an unreduced interaction process within a quantum system described by the ordinary Schrödinger equation, but which is analysed within the universally nonperturbative method of effective dynamical functions (see the previous section) that gives redundant system realisations and the ensuing causal randomness permitting one to understand the fundamental origin of the true quantum chaos in a closed, Hamiltonian system (absent in the conventional approaches), in its agreement with the usual principle of correspondence, and the causal, dynamic origin of quantum uncertainty and "wave reduction (collapse)" in the process of quantum measurement considered as interaction within a totally quantum, but very slightly dissipative (open) system of "object" and "instrument".

Although the latter result provides the causal solution of the old "quantum mysteries" (quantum uncertainty and wave reduction) related to the quantum measurement process, the whole problem of (causally complete) foundation of quantum (wave) mechanics does not stop there, since one should also understand, and rigorously describe, the genuine physical nature of a free "elementary particle", including its "corpuscular" and "undulatory" properties, their intrinsic duality, the causally complete meaning of the "wavefunction", "quantization", other related features, and finally propose a consistent *derivation* of the Schrödinger equation and the accompanying canonical "postulates", such as "Born's probability rule". All these results, as well as many others, have indeed been obtained and generalised to any level of dynamic (complex) behaviour within the same dynamic redundance paradigm, proving explicitly its universality [11-14], but the critically important starting point for the whole picture is the definition of the initial "system with interaction", the simplest "starting configuration" of the world, giving rise to its (free) elementary particles. The criterion that should govern the choice of fundamental interaction, giving rise eventually to all observed structures of the world, is the demand for its simplest possible configuration involving practically no artificial entities, "postulates", "principles", or "rules", since even the most fundamental of them should causally *result* from the natural process of world structure formation considered here, by definition, from the really "first principles".

Since there are two, and only two, universal, omnipresent, and "extended" real entities in the observed dynamical structure of the world, the electromagnetic (e/m) and gravitational fields/phenomena, it is natural to assume — and this assumption constitutes the only one, unavoidable and reality-based, "postulate" of the universal science of complexity — that the simplest system with interaction giving rise to all other structures of the world is represented by the a priori absolutely homogeneous system of two interacting, physically real "protofields" of the e/m and gravitational origin, respectively, which are (homogeneously) attracted to one another. Since the protofields are physically real entities, it is clear that they possess a finite compressibility which eventually gives rise to "forces of reaction" compensating their attraction in the situation of "maximal approach" (compression); these forces of repulsion are related to the necessarily existing fine structure of the protofields, but the details of this structure cannot be perceived within this world, since its smallest emerging elements are larger than the protofield elements, and therefore we can consider the protofields to be practically homogeneous. This large enough separation between the successive structural levels of being is a consequence of intrinsic discreteness (quantization) of complex dynamics, causally resulting from the unreduced, holistic character of the driving interaction process [11-14].

We obtain thus indeed the simplest imaginable system of only two interacting entities without any additional imposed structure. Structure appears as a result of unreduced development of attractive interaction between protofields and is observed in the form of "waves" and "particles" of the world (and further emerging more involved structures), whereas the "unperturbed", "free" protofields cannot be directly observed in this world, but remain nevertheless physically real and provide causal extension of the classical notion of "ether" [11-14], superficially rejected by the mechanistic, purely formal approach of the "new physics". As a matter of fact, the observed world structure is asymmetrically "displaced" towards its e/m component, which is related to somewhat different "mechanical" properties of the protofields, where the e/m protofield plays the role of a "rapid" ("light") and nondissipative component that quickly "conforms" to the relatively "slow" ("heavy") and dissipative ("viscous") dynamics of the gravitational medium, so that any structure of the latter is almost immediately "covered" with an e/m "coating". Note that it is still absolutely necessary to explicitly *create* (or obtain as a result of some previous development) this simplest system with interaction in order it can then autonomously produce the observed world structure (contrary to the dominating ideas of the unitary cosmology, formally "justified" by the mechanistic "energy conservation law", about the possibility of universe emergence "from nothing"). On the other hand, if the protofield attraction with generic parameters is considered within the canonical perturbative (and therefore single-valued) analysis, then one obtains a single, more or less homogeneous, "fall" of the protofields onto one another, without further development of structure that could resemble the observed universe. The unreduced interaction analysis, leading to the universal concept of complexity and briefly reviewed below (see refs. [11-14] for more details), gives a qualitatively different result, the chaotic "quantum beat" dynamics providing causal "wave-particle duality" of the emerging elementary particle, causally complete interpretation of the "wavefunction", inertial (and gravitational) mass, and physically real "de Broglie wave" of a moving particle, provided with the intrinsic, causally extended relativity. Further autonomous development of the same interaction process towards higher levels of complexity (structure formation) proceeds by the same mechanism of dynamical splitting into multiple incompatible realisations and gives all existing structures and phenomena including the four "fundamental" interaction forces between particles (intrinsically unified from the beginning), quantum chaos and measurement, the naturally emerging "classical" (trajectorial) behaviour in a closed system, and so on, up to the highest forms of human brain activity [11].

We express the protofield interaction by the *existence equation* that simply fixes the fact of existence of a compound system consisting from the interaction participants, while the details of their resulting (physical) entanglement should be specified within further (unreduced) analysis:

$$\left[ h_\text{e}(q) + V_\text{eg}(q,\xi) + h_\text{g}(\xi) \right] \Psi(q,\xi) = E \Psi(q,\xi), \tag{3}$$

where $q$ and $\xi$ are continuous (but maybe also discretely structured), physically real degrees of freedom of the e/m and gravitational protofields/media, respectively, $h_\text{e}(q)$ and $h_\text{g}(\xi)$ are the corresponding "generalised Hamiltonians" (or any other suitable functions) describing the (unobservable) free-state dynamics of protofields without interaction, $V_\text{eg}(q,\xi)$ is the (eventually attractive) interaction potential, $\Psi(q,\xi)$ is the "state-function" describing the (developing) compound system state, and $E$ is the "eigenvalue" characterising the property expressed by the "generalised Hamiltonians" in this state (as the following analysis shows, it is always reduced to a measure of dynamic complexity, expressed by the generalised energy in the resulting standard description [11]). We emphasize that eq. (3) does not involve any additional assumption, apart from the fact of compound system existence, and practically any known particular equation can be presented in this or equivalent form. Note also the absence of time and usual space in this starting equation; the "degrees of freedom", $q$ and $\xi$, represent the matter of (unperturbed) protofields (or "ether") that cannot be directly perceived in this world. Insisting on the "first principles", "emergent" (and thus totally consistent) type of description, we "have no right" to specify, at this stage, any particular interaction configuration or other details: all the details should self-consistently, dynamically "emerge" in the course of interaction development exactly reproduced by the unreduced description. An important "by-product" of such unlimited generality is the absolutely universal character of the obtained results that can be further applied at any higher level of the world structure emergence, while the difference between the appearing levels is determined by (growing) dynamic complexity (in the form of generalised entropy) represented by the number of the explicitly obtained realisations or rate of their change (see the previous section).

As we have explained above (section 2.1), the unreduced interaction process development leads to formation of multiple, incompatible states of the compound system, called realisations, which are represented, in the case of interacting protofields, by locally squeezed (dynamically compressed, or "reduced") states of closely entangled protofields, centred around different "points" of the thus emerging, *physical* space. This dynamic redundancy of equally possible reduction centres necessarily leads to their permanent, *causally random* change by the system that performs permanent nonlinear, spatially chaotic "oscillation"/"jumps" between successive reduction centres always passing by the "extended", "decompressed" phase of transiently disentangled protofields (before they entangle around the next reduction centre).

This physical picture results from the unreduced mathematical analysis of the starting existence equation within the method of effective dynamical functions [45] (generalising the usual "effective potential method") which will not be reproduced here in detail (see refs. [11-14,46,47]). The causally complete general solution of eq. (3) is provided by the *intrinsically probabilistic* sum of the observed (generalised) system densities, $\rho(q,\xi)$, for each of the obtained incompatible solution-realisations (numbered by index *r*) from the complete set of $N_\Re$ realisations:

$$\rho(\xi,Q) = \sum_{r=1}^{N_\Re} {}^\oplus \rho_r(\xi,Q), \tag{4}$$

where each *r*-th realisation is endowed with the *dynamically derived* probability $\alpha_r$ (cf. eq. (2)), so that the "expectation value" of $\rho(q,\xi)$, obtained in a long enough (or repeated) observation over the system, is given by

$$\rho_{\text{ex}}(q,\xi) = \sum_{r=1}^{N_{\Re}} \alpha_r \rho_r(q,\xi) \ . \tag{5}$$

The detailed solution is specified by the following expressions of the "effective" formalism:

$$\rho_r(q,\xi) = |\Psi_r(q,\xi)|^2 \ , \tag{6}$$

$$\Psi_r(q,\xi) = \sum_i c_i^r \left[ \phi_0(q) + \sum_n \phi_n(q) \hat{g}_{ni}^r(\xi) \right] \psi_{0n}^r(\xi) \ , \tag{7a}$$

$$\psi_{ni}^r(\xi) = \hat{g}_{ni}^r(\xi)\psi_{0i}^r(\xi) \equiv \int_{\Omega_\xi} d\xi' g_{ni}^r(\xi,\xi')\psi_{0i}^r(\xi'), \quad g_{ni}^r(\xi,\xi') = V_{n0}(\xi') \sum_{i'} \frac{\psi_{ni'}^0(\xi)\psi_{ni'}^{0*}(\xi')}{\eta_i^r - \eta_{ni'}^0 - \varepsilon_{n0}} \ , \tag{7b}$$

where $\{\phi_0(q),\phi_n(q)\}$ are the eigenfunctions and $\{\varepsilon_0,\varepsilon_n\}$ eigenvalues of the "free" e/m protofield (generalised Hamiltonian $h_e(q)$), constituting a complete set and actually chosen in the form of δ-like, localised functions representing the irresolvable structural elements of the e/m protofield (while $\varepsilon_0, \varepsilon_n$ correspond to "coordinates of maxima" of those functions), $\varepsilon_{n0} \equiv \varepsilon_n - \varepsilon_0$, $n \neq 0$ takes integer values;[5]

$$V_{nn'}(\xi) = \int_{\Omega_q} dq \phi_n^*(q) V_{\text{eg}}(q,\xi) \phi_{n'}(q) \ ; \tag{8}$$

$\{\psi_{0n}^r(\xi), \eta_i^r\}$ is the complete set of eigenfunctions and corresponding eigenvalues of the *effective existence equation*, numbered by index $i$ within each realisation (the redundant realisations numbered by $r$ are obtained just from this effective equation):

$$\left[ h_g(\xi) + V_{\text{eff}}(\xi;\eta) \right] \psi_0(\xi) = \eta \psi_0(\xi), \tag{9}$$

with the operator of the *effective (interaction) potential (EP)*, $V_{\text{eff}}(\xi;\eta)$, given by

$$V_{\text{eff}}(\xi;\eta) = V_{00}(\xi) + \hat{V}(\xi;\eta), \quad \hat{V}(\xi;\eta)\psi_0(\xi) = \int_{\Omega_\xi} d\xi' V(\xi,\xi';\eta)\psi_0(\xi'), \tag{10a}$$

$$V(\xi,\xi';\eta) = \sum_{n,i} \frac{V_{0n}(\xi)\psi_{ni}^0(\xi)V_{n0}(\xi')\psi_{ni}^{0*}(\xi')}{\eta - \eta_{ni}^0 - \varepsilon_{n0}} \ , \tag{10b}$$

so that its action on an already found eigenfunction, $\psi_{0n}^r(\xi)$, is expressed as

$$V_{\text{eff}}(\xi;\eta_i^r)\psi_{0i}^r(\xi) = V_{00}(\xi)\psi_{0i}^r(\xi) + \sum_{n,i'} \frac{V_{0n}(\xi)\psi_{ni'}^0(\xi)\int_{\Omega_\xi} d\xi' \psi_{ni'}^{0*}(\xi')V_{n0}(\xi')\psi_{0i}^r(\xi')}{\eta_i^r - \eta_{ni'}^0 - \varepsilon_{n0}} \ ; \tag{10c}$$

---

[5] Note that "matrix elements" used in eqs. (7)-(11) and presented here in the canonical form of "overlap integral", eq. (8), can have other expressions with similar role in other problems; variation of such details cannot change the essential result.

and finally $\{\psi_{ni}^0(\xi),\eta_{ni}^0\}$ are the eigenfunctions and eigenvalues of a "truncated" problem characterising other levels of (fractal) complexity development and represented by the "auxiliary" system of equations:

$$\left[h_\mathrm{g}(\xi)+V_{nn}(\xi)\right]\psi_n(\xi)+\sum_{n'\neq n}V_{nn'}(\xi)\psi_{n'}(\xi)=\eta_n\psi_n(\xi). \tag{11}$$

The coefficients $c_i^r$ in eq. (7a) are obtained from the "dynamical" boundary (initial) conditions attained in the transient phase of disentangled, quasi-free interaction participants during their chaotic "jump" from one localised realisation to another. They actually determine realisation probabilities $\{\alpha_r\}$ [11], which can be understood physically: the protofield reduction probability is higher for those realisations (centres of reduction) where the average of chaotically changing protofield magnitude is higher (the probabilities are proportional to squared modulus of the field amplitude expressed by the state-function). It is not difficult to understand that the transitional phase of the system, forming a special, "intermediate", or "main", realisation of the system, provides the causal, *physically real extension of the canonical "wavefunction"*. It becomes clear that the above relation between the irregularly changing, delocalised wavefunction magnitude and the probability of emergence of the highly localised, corpuscular state-realisation, obtained by the physically real wavefunction reduction, provides the causally complete substantiation of the "Born rule of probabilities" stating that the measured probability of particle emergence (event) is proportional to the squared modulus of the wavefunction (due to universality of our description, this causally substantiated rule is reproduced at all higher levels of interaction process development). Therefore in practice one uses the equation for the wavefunction at a higher sublevel of complexity, such as Schrödinger or Dirac equation, that can be causally derived within the universal science of complexity [11-14], in order to determine the distribution of probabilities of localised (corpuscular) state emergence (cf. eq. (2)).

We have provided the main mathematical expressions of the unreduced interaction analysis in order to demonstrate that the obtained qualitatively new results have not only a transparent physical interpretation, but also rigorous formal substantiation (contrary to the canonical fundamental science always obliged to use a number of formally imposed, "mysterious" postulates, quickly growing with the level of system complexity). In particular, the key phenomenon of physically real, dynamical squeeze, or "reduction" ("collapse"), of a portion of extended interacting protofields is a manifestation of the omnipresent instability of any unreduced interaction, completely rejected by conventional, perturbative approaches and originating, within our analysis, in the EP dependence on the eigenvalue to be found, $\eta$, eqs. (9)-(10). This dependence, arbitrarily "eliminated for convenience" in any scholar formalism (e. g. [42,43]), describes the self-consistent local "amplification of interaction" by dynamically emerging "feedback loops": a change of $\eta$ at the right-hand side of eq. (9) will provoke the EP change at the left-hand side, which will further influence the $\eta$ value on the right, etc. The physical basis of this self-amplified instability is also quite transparent: if around certain location an occasional slight approach between the formally homogeneous protofields occur, then their attraction around that location should also slightly increase which will provoke further local approach/squeeze, etc., until the state of maximal compression, where the opposite "forces of reaction" compensate the attraction. The localised, or "corpuscular", character of the resulting state-realisation is described by the resonant terms in eqs. (7b) and (10c), among which only a small, concentrated group has sufficiently high magnitude due to the cutting action of integrals in the numerators. It is important that the same dependencies and related instability mechanism occur for both state-function, eq. (7b), and EP, eq. (10c), which means that the state-function dynamically squeezes around certain centre because the EP self-consistently forms a "potential well" (which *was not* there before and *shall disappear* after realisation emergence) just around the same, randomly chosen location. Causal randomness of this choice of the current reduction centre is determined mathematically by the same resonant denominators, effectively "counting" the number of possible state-realisations and increasing the maximum power of the eigenvalue to be found ($\eta$) in the characteristic equation for the effective existence equation, eq. (9), to the redundant number of realisations. The corresponding physical aspect of redundance is quite evident from the above instability interpretation: an initial, infinitesimal deviation from "average" homogeneity, determining the location of the resulting reduction centre, cannot be fixed or predicted and occurs in many adjacent places with similar probabilities. In the case of strictly homogeneous general system configuration (isolated elementary particle at rest) the probabilities for all $N_\Re$ realisations will be equal, as it also follows from the causally extended "Born probability rule".

It is not difficult to see that the intrinsic instability of protofield interaction "acts in all directions" and leads to the reverse phase of protofield extension after the self-amplified squeeze stops at the state of maximum compression: adjacent protofield portions start to pull increasingly the "corpuscular" state, which is not amplified any more, to "all sides", which induces protofield disentanglement/extension and the system transiently returns to the delocalised "state of wavefunction", after which a new reduction to a randomly chosen centre occurs, etc. The periodic sequence of reduction events provides the most elementary, inherent "clock" of the world and gives rise to the causally determined *time flow*, which is evidently *unceasing* and naturally *irreversible* because of the causally random (intrinsically probabilistic) choice of consecutive reduction centres that form simultaneously the "tissue" of the emerging, *physically real*, tangible *space* (which resolves also the problem of "configurational space" in the standard quantum mechanics). This elementary space and time, resulting only from the unreduced interaction in the *a priori* homogeneous system, constitute the "embedding", most fundamental entities that serve as a basis for emergence, by the same universal mechanism, of the whole multilevel hierarchy of chaotically switching realisations of the world structure, forming the respective higher levels of space and time, with a specific physical quality of space "tissue" at each level [11] (see section 2.1).

We call the spatially chaotic, time-periodic sequence of cycles of highly nonlinear reduction-extension in the system of interacting protofields *quantum beat*. It represents the *standard* (universal) form of the unreduced interaction development, in the form of chaotic realisation change, and can also be described as chaotic spatial "wandering" of the *virtual soliton*, the latter designating system state in the phase of maximum dynamical squeeze and describing the causal version of the localised, corpuscular state of elementary *particle* as such. The latter has also the dynamically related, *dual* state of extended field, or "wave", so that such *elementary field-particle* is permanently, naturally *transformed*, only due to the same a priori uniform interaction, from one of these *complementary, or dual*, states to another, performing its totally irregular "internal pulsation", equivalent to the *rest mass*, or partially ordered walk, equivalent to global motion (see also below). In this way we obtain the causally complete, intrinsically consistent interpretation of elementary particle, its inherent wave-particle duality, wavefunction, quantization (= dynamical discreteness), and emergent space and time, which brings us much closer to the causal completion of the approach initiated by Louis de Broglie 75 years ago.

In particular, it is evident why and how the above quantum beat process, described by eqs. (7)-(11), realises the causally complete version of the "double solution" by naturally combining within the same, explicitly obtained dynamics both the delocalised "quantum field" and the "singular field" (our "virtual soliton") that were heuristically introduced in the original de Broglie's concept [17-19]. The dynamic transformation between the two components of such extended "double solution with chaos", realising the universal duality of complex dynamics (and providing a realistic extension of ambiguous "complementarity" of Niels Bohr), is evidently possible only due to the dynamic *multivaluedness* phenomenon revealed by the unreduced protofield interaction analysis, which constitutes therefore the "missing link" for the original "nonlinear undulatory mechanics". Causal randomness in the choice of reduction centre, coming from dynamic redundance of interaction process, provides the *physically real* basis for the entity of the wavefunction that is made from the physically tangible "matter", but "behaves strangely" just due to its permanent *chaotic* (intrinsically random) change. The chaotically changing wavefunction serves as a "common blanket" for all emerging "corpuscular" states (virtual soliton locations) and thus provides physical link to the causally complete "hidden thermodynamics" of the virtual soliton, which is none other than the true dynamical chaos of randomly emerging corpuscular

A.P. Kirilyukrealisations of the system, prodigiously predicted by Louis de Broglie [20-23] well before establishment of even its unitary imitations.

Among many intrinsically unified aspects of the obtained "double solution with chaos" [12,13] note especially the qualitatively rich, totally realistic and rigorously justified interpretation of elementary particle as a *complex-dynamical interaction process* (quantum beat) involving already all the intricate details of complex dynamics (intrinsic instability, discreteness, irregularity and unpredictability, fractal entanglement of components).[6] One can compare the obtained picture with the series of purely abstract, and *therefore* "inexplicable", simplifying "postulates" of the usual, unitary theory, leading to artificial manipulations with "state vectors" in purely mathematical "spaces", instead of real objects in real space, in order to understand the crucial advantages provided by the intrinsically realistic de Broglian approach naturally completed by the universal science of complexity. We call application of the latter to the lowest levels of the universal hierarchy of complexity corresponding to "quantum behaviour" of "micro-objects" *quantum field mechanics* [11-14]. It not only eliminates the difficulties of the "nonlinear undulatory mechanics" of Louis de Broglie (including matter wave, double solution, and hidden thermodynamics) with the help of the causally complete concept of true, "internal" dynamical chaos, but leads, as we shall briefly demonstrate below, to its *intrinsic unification with the causally extended relativity* (both "special" and "general") and "field theory"/"particle physics" (involving the dynamically unified "fundamental interactions").

---

*La relation du quantum n'aurait sans doute pas beaucoup de sens si l'énergie pouvait être distribuée d'une façon continue dans l'espace, mais nous venons de voir qu'il n'en est sans doute pas ainsi. On peut donc concevoir que par suite d'une grande loi de la Nature, à chaque morceau d'énergie de masse propre $m_0$ soit lié un phénomène périodique de fréquence telle que l'on ait:*

$$h\nu_0 = mc^2$$

$\nu_0$ *étant mesurée, bien entendu, dans le système lié au morceau d'énergie. Cette hypothèse est la base de notre système: elle vaut, comme toutes les hypothèses, ce que valent les conséquences qu'on en peut déduire.*

*Devons-nous supposer le phénomène périodique localisé à l'intérieur du morceau d'énergie? Cela n'est nullement nécessaire et il résultera du paragraphe III qu'il est sans doute répandu dans une portion étendue de l'espace. D'ailleurs que faudrait-il entendre par intérieur d'un morceau isolé d'énergie? L'électron est pour nous le type du morceau isolé d'énergie, celui que nous croyons, peut-être à tort, le mieux connaître; or d'après les conceptions reçues, l'énergie de l'électron est répandue dans tout l'espace avec une très forte condensation dans une région de très petites dimensions dont les propriétés nous sont d'ailleurs fort mal connues. Ce qui caractérise l'électron comme atome d'énergie, ce n'est pas la petite place qu'il occupe dans l'espace, je répète qu'il l'occupe tout entier, c'est le fait qu'il est insécable, non subdivisible, qu'il forme une unité.*

Louis de Broglie, "Recherches sur la théorie des quanta", Thèse (1924), [5]

---

---

[6] In particular, the unreduced fractal entanglement of the interacting protofields gives rise to the causally complete interpretation of the property of *spin* emerging, due to the inevitable shear instability, as a vortex motion of the e/m protofield "matter" performing its reductive interlacement with the gravitational medium *around* the emerging centre of compressed, corpuscular state [11-14].

Although the obtained picture of "hidden" complex dynamics of the elementary field-particle provides the uniquely consistent, realistic interpretation of the known qualitative patterns of quantum behaviour, it cannot be observed in detail because it represents the lowest level of the world complexity whose internal structure is not resolved in principle (while remaining quite real). What can be observed, however, is the average effect of chaotic quantum beat dynamics of the isolated field-particle, and we argue that the main observed "global" manifestation of the complex field-particle dynamics is none other than its inertial *mass*, universally defined and causally extendible to effects of both special and general relativity (see below). It is not difficult to understand that inertial mass of an isolated particle-process, intrinsically equivalent to its energy, can be consistently interpreted as the temporal rate, or "intensity", of realisation change of the *chaotic* quantum beat process, representing a measure of its *complexity*. Due to the "internal", purely dynamic origin of causal randomness of the virtual soliton wandering "inside" the field-particle, such understanding of the property of mass provides the main feature of *inertia* (resistance to a change of state), which is due to the *already existing* (and irreducible) internal motion and can be compared to resistance to compression of a gas of chaotically moving molecules. In addition, such interpretation of mass provides a natural extension and decisive clarification of de Broglie's ideas about "hidden thermodynamics of isolated particle" and related "variable mass of a particle" [20-23], by finding the necessary source of randomness *within the a priori deterministic* dynamics of a basically simple system and thus permitting one to avoid formal imposition of some external, and inevitably ambiguous, "source of chaoticity", "hidden thermostat", which can produce only "variations" of already existing (and remaining unexplained) "proper" mass.

The mathematical expression of the proposed complex-dynamic definition of elementary particle mass, compatible with usual definitions of energy, $E$, and mechanical action, $\mathcal{A}$, is [11,13,14]:

$$E_0 = m_0 c^2 = -\frac{\Delta \mathcal{A}}{\Delta t} = \frac{h}{\tau_0} = h\nu_0 \;, \tag{12}$$

where index 0 refers to the simplest case of the state of rest, $c^2$ is simply a coefficient for the moment (to be properly specified later as the square of the velocity of light), $-\Delta\mathcal{A} = h$ (Planck's constant) is the quantum of action-complexity corresponding *physically* to one cycle of quantum beat (nonlinear reduction-extension of the coupled protofields or one "quantum jump" of the "corpuscular state", the virtual soliton), $\Delta t = \tau_0$ is the *emerging* "quantum of time" equal to one period, $\tau_0$, of the same cycle, and $\nu_0 = 1/\tau_0$ is the corresponding quantum beat frequency, forming the basis of the causal time concept ($\nu_0 \sim 10^{20}$ Hz for the electron). The obtained relation, $m_0 c^2 = h\nu_0$, is heuristically introduced by de Broglie [1,5] as "inevitable" unification of the main (postulated) "principles" of relativity and quantum theory in the behaviour of a physically real particle and makes the starting point for his causal derivation of the "(matter) wave associated with a (moving) particle". Now this deceptively simple relation is causally substantiated by the above description of the holistic complex-dynamical process within the elementary field-particle, eqs. (7)-(11), so that it is the formal postulates of the standard relativity *and* quantum mechanics that can be consistently derived *together*, in their *natural unity* artificially broken by the canonical science, within the same causally complete picture of complex protofield interaction dynamics. As a result, such quantities as mechanical action $\mathcal{A}$, its natural discrete portion ("quantum") $h$, and the "internal" frequency of a particle ("mobile") acquire quite new, complex-dynamical sense containing the causally specified, dynamic meaning of nonlinearity, instability, and physical protofield entanglement of the quantum beat process [11-14].

Further advantages of this causally complete description of quantum field mechanics appear when we pass to the case of a moving field-particle. We can, first, *objectively* (and universally) define the states of rest and motion, using *only rigorously derived results* of the emergent concept of dynamic complexity that underlies any really occurring process. The *state of rest* of an object is the state with the *minimum* possible dynamic complexity (energy) of that object (this minimum always exists because complexity is a positively defined, finite quantity). It is not difficult to understand that the state of rest thus defined corresponds simply to the most uniform distribution of the system realisation probabilities (it is totally uniform for the isolated field-particle), which means that the quantum beat dynamics of the field-particle at rest is represented by the *totally* irregular wandering of the virtual soliton "within its wavefunction" (any average, global tendency is absent). Any growth of the system complexity-energy above this minimum corresponds to its transition to a *state of motion*, appearing as an average (global) *tendency* in the virtual soliton wandering (its preferred orientation within the thus emerging "direction of motion"), whereas each individual "jump" of the virtual soliton (i. e. system realisation change) *always* preserves its irreducibly *probabilistic*, unpredictable character. Therefore a state of motion is characterised by a certain *order* (inhomogeneity) in the *distribution of realisation probabilities* that accounts eventually for *all* observed structures of the world and is associated with the corresponding growth of complexity-entropy [11] (see also section 2.1).

Global order in virtual soliton wandering means that a state of motion of the field-particle appears as an observable *spatial* degree of freedom of its wave field, oriented in its "direction of (global) motion" and obtained due to increase of its mass-energy-complexity over the minimum of the state of rest (eq. (12)), so that the relation of eq. (12) is transformed, for the moving field-particle, into

$$E = mc^2 = -\frac{\Delta \mathcal{A}}{\Delta t} + \frac{\Delta \mathcal{A}}{\lambda}\frac{\Delta x}{\Delta t} = \frac{h}{T} + \frac{h}{\lambda}v = hN + pv \; , \tag{13}$$

where

$$E = -\frac{\Delta \mathcal{A}}{\Delta t}\Big|_{x = \text{const}} = \frac{h}{\tau} = h\nu \; , \tag{14}$$

$$p = \frac{\Delta \mathcal{A}}{\Delta x}\Big|_{t = \text{const}} = \frac{\Delta \mathcal{A}}{\lambda} = \frac{h}{\lambda} \; , \tag{15}$$

$$v = \frac{\Delta x}{\Delta t} \equiv \frac{\Lambda}{T} \; , \tag{16}$$

$\lambda \equiv (\Delta x)|_{t = \text{const}}$ is the *emerging* "quantum of space", an elementary directly measurable (regular) space inhomogeneity characterising the elementary field-particle with complexity-energy $E$ ($> E_0$) and resulting from its global displacement (motion), $\Delta t = T$ is the "total" period of nonlinear quantum beat in the state of motion with complexity-energy $E$ ($N = 1/T$ is the corresponding quantum-beat frequency), $\Delta x = \Lambda$ is the "total" quantum of space, while $\tau \equiv (\Delta t)|_{x = \text{const}}$ is the quantum-beat period measured at a fixed space point (we thus obtain here the naturally quantized and realistic versions of the canonical "partial" and "total" derivatives and corresponding increments).

It is not difficult to understand that the elementary space inhomogeneity $\lambda$ of the wave field of a moving field-particle, thus *dynamically* emerging in the underlying process of quantum beat, is none other than "de Broglie wavelength of a particle", $\lambda \equiv \lambda_B = h/p$, providing this rather ambiguous notion of the scholar quantum theory with a causally complete interpretation directly completing its meaning introduced by Louis de Broglie in his thesis [5], but then superficially neglected by the unitary, postulated science. The global motion tendency of the field-particle, corresponding to its increased energy-complexity and described by the second summand in eqs. (13), appears as a real, "corrugated" (or "undulatory") structure in the physically real wave field within the e/m protofield always coupled to its (hidden) gravitational partner. The spatial period of undulation equals to $\lambda_B$, but despite its canonical presentation in the form of a linear, "plane" wave, this undulatory structure has essentially nonlinear, complex-dynamical nature, just providing unified causal explanation for the associated "mysterious" peculiarities of quantum (and in reality complex-dynamic) behaviour. This global "corrugation" with the period of $\lambda_B$ has a dynamically changing, but at the same time "fixed", structure of a standing-travelling, nonlinear "wave", which is indeed fixed in its nodes, but permanently makes "occasional" (irregular in time), quantized displacements to the distance of $\lambda_B$ (in reality, it is always the virtual soliton that "produces" the undulation and performs average displacement/motion, superimposed on the background of its totally irregular wandering described by the first summand in eqs. (13)). In general, the emergence of the undulatory space inhomogeneity in the wave field of a moving field-particle can be compared to appearance of waves at the homogeneous sea surface under the (uniform) influence of wind, where the gravitational (proto)field influence on the moving "matter" plays an essential role in both cases, but in the case of elementary field-particle all the effects and motions are extremely nonlinear and necessarily quantized into indivisible quantum beat cycles, each of them corresponding to the increment of $-h$ of the (extended) action-complexity.

In addition to the obtained causally complete interpretation of "matter wave" and the related "wavefunction", the quantum beat dynamics provides the causal interpretation of *relativistic effects*, considerably extending their formal, mechanistic imposition by the canonical "postulates" and artificially adjusted "principles". Thus, the above intrinsic combination of irregular and global (regular) tendencies in the same causally probabilistic process provides physically complete explanation for the equivalence between mass and energy, as well as causal derivation of the "relativistic dispersion relation" [13,14],

$$p = E\frac{v}{c^2} = mv, \qquad (17)$$

where $m \equiv E/c^2$, now by rigorously substantiated definition. The physical sense of this deceptively simple relation is that it establishes the connection between the two tendencies within the same field-particle dynamics (regular global motion and irregular deviations from it): the explicitly observed global (regular) tendency represented by momentum ($p$) constitutes only a part (the proportion of $\beta = v/c$) of the whole dynamics represented by the total energy, $E$. This causal substantiation of the "familiar" (but always postulated) relation of eq. (17) has many important consequences. In particular, Newton's laws, now in their causally complete (and relativistically extended) form, can be obtained from eq. (17) by taking its time derivative. Substitution of eq. (17) into expression for $\lambda$ of eq. (15) gives the detailed expression for the de Broglie wavelength,

$$\lambda = \lambda_B = \frac{h}{mv}, \qquad (18)$$

which coincides with the canonical expression, eq. (1), but implicitly includes the nontrivial complex dynamics of quantum beat providing the causally complete solution of all contradictions existing around this relation. Thus, one can avoid using any fictitious "wave of phase" moving with the (variable) superluminal velocity and giving the actual particle velocity $v$ only as the group velocity [5], which contradicts, in particular, to the nonlinear character of the causal wave mechanics. The physical picture of complex dynamics behind the "dispersion relation" of eq. (17) shows that in reality the part of the total energy leading to "superluminal" motion of the regular matter wave in the original interpretation is dispersed in the "hidden", but irreducible, irregular wandering of the virtual soliton "around" the observed regular wave, so that the remaining part just gives the real, internally nonlinear matter wave travelling with the particle velocity $v$ (though in a quantized, "quasi-standing" mode described above). This interpretation demonstrates again that the true dynamical chaos, based on the dynamic redundance phenomenon, forms just the necessary "missing link" for the original de Broglie's theory and perfectly completes it to the totally consistent and adequate description of quantum (and relativistic) behaviour.

Using the dispersion relation of eq. (17) in the complex-dynamic partition of the total energy, eq. (13), we obtain the relation between different time measures, $\tau$ and $T$, of quantum beat period:

$$\tau = T\left(1 - \frac{v^2}{c^2}\right). \qquad (19)$$

This expression can be further transformed to a more convenient form if we use an additional relation between the time periods which also follows from the underlying complex dynamics [11,13,14]:

A.P. Kirilyuk

$$N\nu = (\nu_0)^2, \tag{20a}$$

or

$$T\tau = (\tau_0)^2. \tag{20b}$$

Combining eqs. (19) and (20), we obtain the following relations:

$$T = \frac{\tau_0}{\sqrt{1 - \frac{v^2}{c^2}}} \quad \text{or} \quad N = \nu_0 \sqrt{1 - \frac{v^2}{c^2}}, \tag{21}$$

$$\tau = \tau_0 \sqrt{1 - \frac{v^2}{c^2}} \quad \text{or} \quad \nu = \frac{\nu_0}{\sqrt{1 - \frac{v^2}{c^2}}}. \tag{22}$$

Since, according to the above results, the quantum beat periods $\tau_0$ and $T$ determine the causal flow of "internal" system time in the state of rest and motion, respectively, the relations of eq. (21) reproduce the famous "relativistic time retardation" effect, now provided, however, with the physically complete understanding of the underlying, irreducibly complex dynamics. We see that the true origin of "time relativity", i. e. its dependence on motion, is not in a postulated abstract "principle" about "laws of nature" and their mathematical expression (the standard, Einsteinian interpretation of Lorentz-Poincaré relations), but in the fact that the real *time itself* is causally *"produced"* within the *same* complex-dynamical process that gives global *motion* of a particle and eventually any larger "body" consisting from many particles. When a particle (or body) is set in motion, all the participating elementary quantum beat processes should accelerate their reduction-extension frequency, but especially its contribution to the regular, global tendency, whereas the remaining irregular tendency, constituting the whole dynamics of the body *at rest* and *always* determining the flow of the "internal" system time, should inevitably decrease its magnitude, and thus frequency, which is observed as relative retardation of the "internal" time. This physically complete interpretation provides decisive confirmation for the evident general argument: one can obtain a really consistent interpretation of relation between time and motion (or "time relativity") only by specifying the physical origin of this relation, inevitably involving the causal *origin of time*. Needless to say, this was not and could not be done within the canonical, linear and superficially formalised "new physics", whatever are its external, pseudo-philosophical pretensions.

Other relations from the canonical "special relativity" can also be easily derived [11,13,14], in the same causally complete way, involving no additional postulates and imposed "principles" about a "form of the Lagrangian", etc. Thus, combining eqs. (22) with the causally derived energy-time and energy-mass relations from eqs. (12)-(14), (17), one obtains the expression of "relativistic increase of mass" of a moving field-particle (and a many-particle body):

$$m = \frac{E}{c^2} = \frac{m_0}{\sqrt{1 - \frac{v^2}{c^2}}}. \tag{23}$$

Let us emphasize once more the inseparable unity of "quantum" and "relativistic" manifestations of the unreduced dynamic complexity of the same protofield interaction process, as opposed to artificial "joining" of "quantum" and "relativistic" postulates in the canonical, unitary science. Moreover, universality of the dynamic redundance paradigm permits us to generalise both "quantum" and "relativistic" aspects of complexity to any its higher, "classical" level [11], which demonstrates once more their causal, realistic origin, escaping the "mystified" approach of the canonical "new physics", which is closely related to huge simplification of being within the inherent one-dimensional thinking.

The complex-dynamical partition of the total field-particle energy, eq. (13), that was emphasised by de Broglie within his "hidden thermodynamics" concept, can now be rewritten in the most complete form reflecting the obtained causal interpretation of the related quantum and relativistic features of complex dynamics:

$$E = h\nu_0\sqrt{1-\frac{v^2}{c^2}} + \frac{h}{\lambda_B}v = h\nu_0\sqrt{1-\frac{v^2}{c^2}} + h\nu_B = m_0c^2\sqrt{1-\frac{v^2}{c^2}} + \frac{m_0v^2}{\sqrt{1-\frac{v^2}{c^2}}} , \qquad (24)$$

where $h\nu_0 = m_0c^2$ (eq. (12)) and we introduce *de Broglie frequency*, $\nu_B$, defined as

$$\nu_B = \frac{v}{\lambda_B} = \frac{pv}{h} = \frac{\nu_{B0}}{\sqrt{1-\frac{v^2}{c^2}}} = \nu\frac{v^2}{c^2} , \quad \nu_{B0} = \frac{m_0v^2}{h} = \nu_0\frac{v^2}{c^2} = \frac{v}{\lambda_{B0}} , \quad \lambda_{B0} = \frac{h}{m_0v} . \qquad (25)$$

Note that the direct relation with de Broglie's "thermodynamical" expression of participating quantities can also be established [11].

We can now only very briefly outline further essential results of the quantum field mechanics, some of them going far beyond the causal "matter wave" as such, but remaining always intimately related with the same irreducibly realistic approach that was maintained by Louis de Broglie who always insisted on realistic "description in terms of space and time" and the largest possible unity of the obtained understanding of reality.

Causal substantiation of the canonical "quantization rules" is based on the described complex-dynamic interpretations of the wavefunction, wave-particle duality, space and time, energy-mass and momentum. It leads to the causally derived wave equations, such as Schrödinger and Dirac equations, clearly explained now as expressions of the same irreducible complexity of underlying quantum beat processes and their development towards higher levels of complexity (structure formation) [11,13,14]. The famous "uncertainty relations" and "energy level discreteness" emerge as natural, "standard" manifestations of the dynamical discreteness (quantization) and duality of a real, necessarily complex interaction process.

All the basic, "inherent" properties of elementary particles, such as mass, spin, electric charge, "type of statistics" ("bosons and fermions"), as well as related quantized interactions (see also below), obtain their causally complete, physically transparent and detailed interpretation, liberated from any canonical "mysteries", as manifestations of the same intrinsic dynamic complexity of the quantum beat processes within (and between) field-particles, similar to the case of inertial (relativistic) mass-energy described above in more detail. The canonical, empirically justified "conservation laws" of the corresponding quantities (mass-energy, momentum, charge, etc.) are naturally unified now within the single, causally substantiated and absolutely universal law of conservation (symmetry) of complexity that provides in addition the universal law of system evolution (see also section 2.1).

The "classical", permanently localised (trajectorial) behaviour is obtained as a particular level of complex behaviour, generally coinciding with formation of elementary bound systems (such as atoms), simply because the probability of *correlated random* jumps of several closely connected virtual solitons to a larger distance is low and quickly (exponentially) drops with distance. Note that due to the dynamic reduction and true chaoticity of the *internal* particle dynamics, no ambiguous "external influence" producing (mathematical) "decoherence" of (abstract) "state vectors" is necessary for emergence of the *physically localised* behaviour in our approach, contrary to imitative explanations of classicality within various unitary "interpretations" (see e. g. [30]).

Another idea proposed by Louis de Broglie in connection with his "hidden thermodynamics" concept finds its confirmation and natural completion within the quantum field mechanics. It is a causal extension of the canonical "least action principle" that deals with fictitious trajectories of a dynamical system, among which it should actually choose one, "real" trajectory, according to the canonical "variational principle". In agreement with the corresponding suggestion of de Broglie, it appears that in reality those "wandering" trajectories are not fictitious, but quite *real* ones, and correspond to realisations chaotically taken by the system with generally different probabilities, so that an average, more or less distinct, "real" trajectory of the canonical science can emerge as the average tendency, providing a better or worse approximation for the detailed, irreducibly chaotic (dynamically multivalued) system path. This leads also to causal modification and generalisation of "Feynman path integrals" also dealing, in their canonical version, only with fictitious "trajectories" of abstract "state vectors". The underlying universal principle is *conservation (symmetry)* of the total dynamic complexity realised by unavoidable growth of complexity-entropy ("extended entropy growth law") at the expense of equal decrease of complexity-action tending to its smaller values ("modified least action principle").

Gravitational attraction between particles (and bodies), as well as the effects of "general relativity" (such as "time retardation in the gravitational field") and equivalence between inertial and gravitational masses, appear within the same unified picture of interacting protofields due to the physical transmission of the influence of quantum beat reductions of each particle-process on the (local) tension of gravitational medium (protofield) to all other particles obtained in the same medium interaction. Note the fundamental difference of such understanding of gravity as a *complex-dynamical process* (including *dynamic quantization*) from the purely abstract, "geometric" interpretation within the canonical Einsteinian relativity; both formal "unification" of abstract space-time within a single "manifold" and any ambiguous "deformation" of this imaginary construction are rejected by the quantum field mechanics, which attributes gravitational effects to inhomogeneous, and dynamically changing, distribution of real "mechanical" properties (tension, density) within the real gravitational medium/protofield. It is important that dynamic, gravitational (including "relativistic"), and quantum effects emerge in our approach in their intrinsically unified form of the unreduced complex dynamics.

This natural unifying tendency of the quantum field mechanics continues towards causal interpretation of *dynamically unified* "fundamental interaction forces" between particles, the result that has never been attained by the canonical "field theory", despite much efforts devoted to the abstract unification of simplified formal schemes, basically detached from reality, but provided with some arbitrary, speculative interpretations. Namely, whereas gravitational interaction is dynamically transmitted, in the described way, through the gravitational medium, the ordinary e/m interaction is transmitted, in a generally similar, physically real way, through the e/m protofield; both interactions are naturally quantized due the quantized character of the quantum beat dynamics. The "strong" and "weak" interactions are equally consistently interpreted as extreme short-range ("contact") interactions between the constituent elements of the gravitational and e/m medium, respectively, dynamically unified with other two (long-range) interactions through the common, physically real "carriers" (the two protofields) and their interaction, especially in the phase of maximal compression (wave reduction) of the quantum beat cycle. Partial unifications between e/m and weak interactions, on one hand, and gravitational and strong interactions, on another hand (this latter one constituting the new result), are also clearly explained as being due to the common physical "carrier"-protofield within each couple. Similarly, the total number, 4, of fundamental interactions acquires most natural explanation as being due to the world construction from two protofields multiplied by two characteristic, sufficiently separated ranges, long and short, of interaction transmission within each protofield: $2 \times 2 = 4$, as it could be expected, and the fact that over-sophisticated mechanistic constructions of the canonical "field theory" ("strings", "branes", "supersymmetry", "great unifications", etc.) can never attain this result demonstrates that the true, dynamic complexity of nature is always hidden behind the elegant simplicity of its external manifestations, and not the reverse!

We can clearly see now that the externally simple expression of the undulatory mechanics of a particle, eq. (1), proposed by Louis de Broglie 75 years ago, contains the unique combination of practically all fundamental aspects of reality intrinsically unified by the unreduced dynamic complexity of the quantum beat process, as it was prodigiously predicted within the original de Broglie approach. This natural unity of complexity, inertia, energy, gravitation, quanta, waves, particles, their duality, and relativity is even *directly* expressed (but never recognised) within that remarkably compact formulation, $h = \lambda_B m v$, which can now be translated into the profound physical, causally complete relation between the emerging, dynamically connected aspects of the unreduced, complex-dynamic reality:

$$\left\{ \begin{array}{c} h \\ quantized \\ complexity \end{array} \right\} = \left\{ \begin{array}{c} \lambda_B \\ wave\ behaviour \\ undulatory\ motion\ structure \end{array} \right\} \otimes \left\{ \begin{array}{c} m \\ relativistic\ inertial\ and \\ gravitational\ mass/energy \end{array} \right\} \otimes \left\{ \begin{array}{c} v \\ corpuscular\ behaviour \\ trajectorial\ space\ structure \end{array} \right\}.$$

Causal consistency of quantum field mechanics is confirmed by the growing number of its practically important consequences. One of them appears in the form of renormalised, causally derived Planckian units which, due to their essential difference from the conventional, formally produced values, lead to important new conclusions about the expected mass spectrum of elementary particles and corresponding qualitative change of accelerator research that should be displaced from the vain search for hypothetical, but actually absent "super-heavy" species at ever higher energies to a better oriented investigation of a moderate, already attained energy scale [11,14]. Another practically important application concerns essential extension of the evidently deceptive, but artificially boosted idea of the unitary "quantum computers" to the realistic, chaotic (complex-dynamical) quantum computers [11,14], which needs essential use of the explicitly multivalued basis of the quantum field mechanics instead of canonical, unitary theory. In general, the causally complete understanding of the universal science of complexity provides, by its very origin, the clear orientation, well specified "guiding line" in any particular field of knowledge, fundamental or applied. On the other hand, the canonical, unitary science, devoid of such guiding line and submerged instead into an infinite series of arbitrary speculations and artificial imitations, not only cannot give any qualitatively new knowledge (true scientific discovery), but also will inevitably produce various practical "difficulties" of a catastrophic scale because of its huge, always growing, but completely disoriented and therefore destructive, technical power [14].

### 3. Renaissance thinking against medieval unitarity: The "lost" fight still leads to the victory of unreduced complexity

We shall not describe here further details of the intrinsically holistic, unrestrictedly universal picture of multilevel complex dynamics of the world that progressively emerges within the same dynamic redundance paradigm applied to the unique, *a priori* ultimately simple system of two interacting protofields [11-14]. Note only that the irreducibly realistic approach of the causal science, maintained during the whole twentieth century almost exclusively due to the singular efforts of Louis de Broglie and his followers, always gives, by its very origin, well specified results expressed in terms of consistently reproduced "elements of reality", contrary to the effectively one-dimensional, and therefore fundamentally unreal, paradigm of the canonical, unitary science that has come to its definite end now. The new, ultimately complete and universal knowledge can instead start its development, and the universal science of complexity, largely prepared and "felt" in many details by Louis de Broglie, constitutes a suitable starting point for the forthcoming, inevitable renaissance of the causally complete knowledge.

The fight between mechanistic simplification and unreduced complexity of the real, living world is always won in advance, in favour of reality. And if a decadent civilisation, dominated by intrinsic adherents of low-level practicality, obscurity and perverted mysticism, is unable to find the way to the unreduced reality and take the side of truth, then it will inevitably suffer a series of dramatic falls prepared by the stubborn mediocrity of "chosen" sages within their one-dimensional thinking. Being always an "intellectual refugee" with his singular efforts in favour of the unreduced, causally complete understanding of reality, suffering from quasi-total misunderstanding even in France, his own country for which he had done so much and to which he was so profoundly attached, Louis de Broglie had clearly seen this crucial difference between the objective truth and its mechanistic imitations and took the side of truth without hesitation. Today's situation in quantum mechanics, disgracefully missing the 75th birthday of its true origin (as well as all the passed 75 years), and the crisis in fundamental science in general, confused within it own deceptions, clearly demonstrate that Louis de Broglie, this unrecognised leader of the twentieth century thinking, was right in all his attitudes, and the "grands maîtres" of the "developed" world should better listen to him, at least now.

---

*Il est essentiel de remarquer que, si les formalismes mathématiques permettent seuls dans les sciences où l'on peut les introduire de donner à nos idées une grande précision, ils ne sont pas cependant sans présenter quelques dangers car, entraîné par leur clarté et leur automatisme, on peut facilement oublier qu'ils ne fournissent jamais que les conséquences des hypothèses qui ont été mises à leur base. Seules, l'intuition et l'imagination permettent de briser le cercle dans lequel s'enferme naturellement toute pensée qui veut être purement déductive.*

Louis de Broglie, *Les idées qui me guident dans mes recherches* (1965), In [40]

---

---

*Tandis que, par la force même des choses, s'appesantit sur la recherche et sur l'enseignement scientifiques le poids des structures administratives, des préoccupations financières et la lourde armature des réglementations et des planifications, il est plus indispensable que jamais de préserver la liberté de l'esprit scientifique, la libre initiative des chercheurs originaux parce qu'elles ont toujours été et seront sans doute toujours les sources les plus fécondes des grands progrès de la Science.*

Louis de Broglie, *Nécessité de la liberté dans la recherche scientifique* (1962), In [40]

---